\documentclass[aps,prb,twocolumn,showpacs]{revtex4}

\usepackage{amsmath}
\usepackage{graphicx}
\usepackage{bm}
\usepackage{epsf}

\newcommand \bea {\begin{eqnarray} }
\newcommand \eea {\end{eqnarray}}

\newcommand{\beg}{\begin{equation}}
\newcommand{\en}{\end{equation}}

\begin{document}

\title{Exact solution for quantum dynamics of a periodically-driven two-level-system}

\author{Anirban Gangopadhyay, Maxim Dzero, and Victor Galitski}
\affiliation{Condensed Matter Theory Center and Department of Physics, University of Maryland,
College Park, MD 20742-4111, U.S.A.}

\begin{abstract}
We present a family of exact analytic solutions for non-linear quantum dynamics of a two-level system (TLS) subject to a periodic-in-time external field. In constructing the exactly solvable models, we use a
``reverse engineering'' approach where the form of external perturbation is chosen to preserve an
integrability constraint, which yields a single non-linear differential equation for the ac-field. A solution
to this equation is expressed in terms of Jacobi elliptic functions with three independent parameters that
allows one to choose the frequency, average value, and amplitude of the time-dependent field at will.
This form of the ac-drive is especially relevant to the problem of dynamics of TLS charge defects that cause dielectric losses in superconducting qubits. We apply our exact results to analyze non-linear dielectric response of such TLSs and show that the position of the resonance peak in the spectrum of the relevant correlation function is determined by the quantum-mechanical phase accumulated by the TLS wave-function over a time evolution cycle. It is shown that in the non-linear regime, this resonance frequency may be shifted strongly from the value predicted by the canonical TLS model. We also analyze the ``spin'' survival probability in the regime of strong external drive and recover a coherent destruction of tunneling phenomenon within our family of exact solutions, which manifests itself as a strong suppression of ``spin-flip'' processes and suggests that such non-linear dynamics in LC-resonators may lead to lower losses.
\end{abstract}


\maketitle
\section{Introduction}
The problem of a periodically-driven two-level system (TLS) appears in many physical contexts including
magnetism, superconductivity, structural glasses and quantum information theory.~\cite{Weiss,Leggett1987,AndersonTLS,YuAndersonA15,Hunklinger1981,Peter2002,manipulate} The interest in this old problem has been revived recently due to advances in the field of quantum computing
(see, {\em e.g.}, Refs.~[\onlinecite{Gerardot2009,expJJ1,expJJ2,expJJ3,SlavaScience1}] and references therein).
First of all, a qubit itself is a two-level system and the question of its evolution under an external time-dependent perturbation is obviously of interest. Also, the physical mechanism that currently limits coherence particularly in superconducting qubits is believed to be due to other types of unwanted TLSs within the qubit, whose charge dynamics under a periodic-in-time electric field gives rise to dielectric losses  directly probed in experiment.~\cite{YuPRL,Wang2009} In what follows, we mostly apply our solution to the latter charge TLS model, but the general methods and some particular results of this work evidently can be applied to a much broader range of problems (see, {\em e.g.}, Ref.~[\onlinecite{Grifoni2009}] and references therein).

One of the key metrics of a superconducting qubit is the quality factor, which is defined as a ratio of the
real and imaginary parts of the dielectric response function, $\varepsilon(\omega)$, evaluated at the resonant frequency of the corresponding LC-circuit, $Q =  {\rm Re}\, \varepsilon(\omega_{\rm r})/ {\rm Im}\, \varepsilon(\omega_{\rm r})$. Very high values of the quality factor are required for the qubit to be operational. However, existing experiments consistently show significant dielectric losses that occur in an amorphous dielectric ({\em e.g.}, in Al$_2$ O$_3$) used as a barrier in the Josephson junctions. It is believed that the losses are primarily due to the presence of charge two-level system defects in the barrier and/or the contact interfaces, which respond to an AC electric field in the LC-resonator. It is still unclear what the physical origin of these defects is, but an early work of Phillips~\cite{Phillips1973} as well as very recent comprehensive density functional theory studies of Musgrave~\cite{Charles2010} point to the OH-rotor defects as a very likely source of the dielectric loss. To determine the physical origin and the properties of the TLSs responsible for the dielectric loss is one of the central questions in the field of superconducting quantum computing and it has been largerly the main physical motivation for our work.

\begin{figure}[h]
\includegraphics[scale=0.29,angle=0]{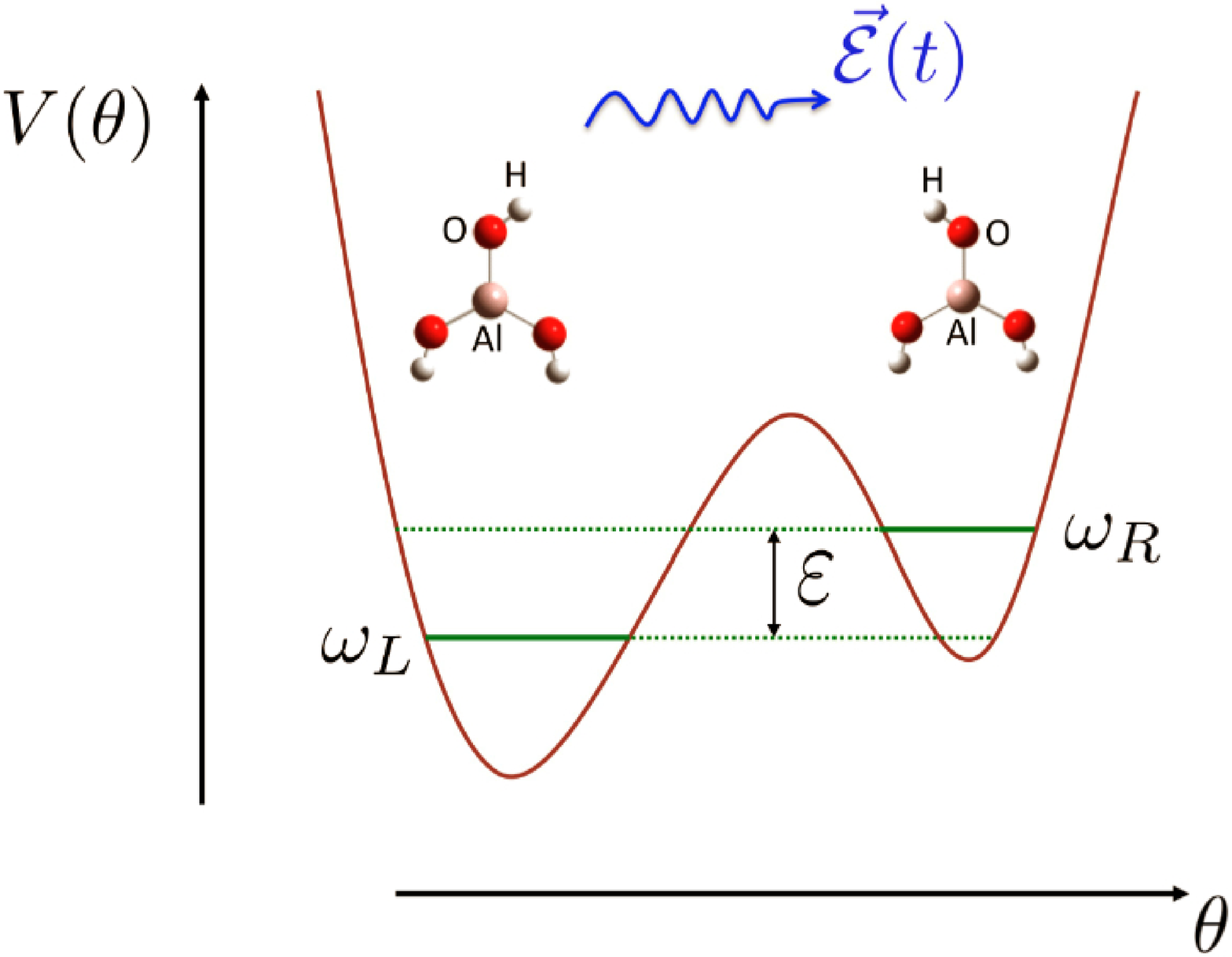}
\caption{Schematic representation of an OH-rotor two-level system  in an Al$_2$O$_3$ oxide.~\cite{Phillips1973,Charles2010}
Here, the role of the generalized variable is assigned to the angle $\theta$ defined as an angle between
the OH-bond and an axis perpendicular to the vertical AlO bond. At low enough temperatures, the
phase space an isolated rotor is reduced to the two-states corresponding to the minima of the double-well potential $V(\theta)$. Application of external ac-field parametrically coupled to the rotor's dipole moment
induces oscillations between the two minima.}
\label{Fig1}
\end{figure}

 The usual theoretical approach to calculating the quality factor and more generally the full dielectric response function, $\varepsilon(\omega)$, involves a formal mapping of charge dynamics in a double-well potential onto the problem of ``spin'' dynamics in an AC field, described by the ``spin'' Hamiltonian $\hat{\cal H}(t) ={\bf b}(t) \cdot \hat{\bm \sigma}/2$, where $\hat{\bm \sigma}$ denotes the Pauli matrices and ${\bf b}(t) =2\left(\Delta_t, 0, \varepsilon + {\vec{d}}_{\rm TLS} \cdot {\vec{E}}(t) \right)$ is an effective ``magnetic field'' that drives TLSs, with  $\varepsilon$, $\Delta_t$, and ${\vec{d}}_{\rm TLS}$  being the TLS energy splitting, the tunneling amplitude between its two states, and the TLS dielectric moment correspondingly and ${\vec{E}}(t)$ is the AC electric field. A linear analysis within the canonical TLS model predicts that the dielectric function due to identical TLSs  is peaked at the frequency, $\nu = \sqrt{\Delta_{\rm t}^2 + \varepsilon^2}$. Ad-hoc inclusion of $T_1$ and $T_2$ relaxation processes and the assumption about random distribution of TLS energy-splittings and tunnelings (typically assumed to be uniform and log-uniform correspondingly) lead to the quality factor $Q  \propto \sqrt{1 + \left(E_0/E_{\rm c}\right)^x}$, with $x\sim 2$, $E_0$ being the amplitude of an applied AC electric field and $E_c$ is a critical value of the amplitude which also encodes
 the information on the strength of the relaxation processes (see, \emph{e.g.}, Ref.~[\onlinecite{Hunklinger1981}]). Both formulas are used widely in interpreting experimental data and probing energetics of the relevant TLS defects \cite{Hunklinger1981,YuPRL}. While this linear analysis  is a fine approximation to describe a majority of regimes currently studied experimentally, the existing experiments are certainly capable and some do~\cite{Osborn2010} access non-linear regimes as well,  where the energy of the applied electric field is comparable or larger than the relevant TLS energies.  Hence, this non-perturbative regime is of clear experimental and theoretical interest. More importantly studies of non-linear dynamics may provide another effective means to probe the properties of TLSs.

The mathematical formulation of the non-linear TLS dynamics problem studied in this paper is deceptively simple: We wish to solve the Schr{\"o}dinger equation for a spinor wave-function, $\Psi = {\psi_+ \choose \psi_-}$, $i\partial_t \Psi = {1 \over 2} {\bf b}(t) \cdot \hat{\bm \sigma}\, \Psi$ that describes a half-integer spin subject to a periodic-in-time magnetic field of the form, ${\bf b}(t) = 2\left( \Delta_{\rm t}, 0 , f(t) \right)$, where $\Delta_{\rm t}$ is a constant describing the coupling between the two states and the function $f(t) = f(t+T)$ describes the time dependent perturbation. Despite the simplicity of the formulation, the problem is generally unsolvable in analytic form for most cases of practical interest. The origin of this surprising fact can be understood if we introduce a new function $R(t) = \psi_+(t)/\psi_-(t)$, which reduces the matrix Schr{\"o}dinger equation to the Riccatti equation $\partial_{(-it)} R =  2 f R + \Delta_{\rm t}  \left[ 1 - R^2 \right]$. It is a non-linear differential equation that has known analytic solution in a very limited number of cases (note that the case of a monochromatic perturbation is not one of them). Therefore, to solve for  TLS dynamics driven by a specific non-equilibrium field is equivalent to generating a particular solution to the Riccatti equation corresponding to this perturbation. Clearly this is a challenging mathematical task and this observation partially explains the current deficit of exact mathematical results.  The difficulties in obtaining exact solutions  have led to the emergence of several perturbative approaches, used in particular to characterize relaxation and dephasing rates in qubits as a function of driving amplitude (see, {\em  e.g.}, Ref.~[\onlinecite{Grifoni2009}] and references therein). These analyses provide very useful physical insights and correctly describe the physics if the time-dependent perturbation is weak, but it is also clear that there  exist non-linear effects beyond perturbation theory and it is desirable to have exact results to access this qualitatively different physics.

The mathematical approach that we use in this paper to obtain exact results is to ``reverse engineer''  exactly solvable Hamiltonians of specific form relevant to the problem of interest. A key observation in our analysis is that finding a Hamiltonian
corresponding to a given solution is much easier than solving the Schr\"{o}dinger equation with a given Hamiltonian. In some generalized sense, the two procedures are related to one another much like differentiation relates to integration. To see this, it is useful to consider the evolution operator, or the $\hat{S}$-matrix, which relates the initial state at $t=0$ to a final state at $t>0$ as follows, $\Psi(t) = \hat{S}(t) \Psi(0)$. In the absence of relaxation process the time-evolution is unitary and it satisfies the Schr{\"o}dinger equation, $i \partial_t \hat{S}(t) = \hat{\cal H}(t) \hat{S}(t)$. If we choose an arbitrary S-matrix, $\hat{S} = \exp{\left[-{i \over 2} {\bm \Phi(t)} \cdot \hat{\bm \sigma}\right]} \in SU(2)_2$, we can immediately reconstruct the corresponding Hamiltonian that gives rise to such evolution as follows $\hat{\cal H}(t) = i \partial_t \hat{S}(t) \hat{S}^\dagger(t)$. Using this method, one can generate an infinite number of exact non-equilibrium solutions and explicit models. These solutions may be of importance to physics of NMR, to the question of  physical implementation of gate operations on a qubit as well as of some mathematical interest. Nevertheless without additional constraints such analyses would generally produce Hamiltonians of little  importance to the problem of dynamics of TLS charge defects.

A very useful insight that allows us to constructively narrow down the range of relevant dynamical systems comes from  the mathematically related problem of far-from-equilibrium superconductivity~\cite{Levitov2004,Emil1,classify}. It is well-known that the reduced BCS Hamiltonian
is algebraically equivalent to an interacting XY-spin model in an effective ``inhomogeneous'' magnetic field in the $z$-direction, whose profile is dictated by the bare single particle-energy dispersion. Far from equilibrium, dynamics of a given Anderson pseudospin \cite{Anderson1958} is determined by an effective time-dependent self-consistent field of other pseudo-spins that it  interacts with.~\cite{VG2010} In many cases (determined by specific initial conditions), this BCS self-consistency constraint dynamically selects a specific order-parameter, such that the dynamics of essentially infinite number of spins is equivalent to the dynamics of few spins only.~\cite{classify} For special sets of initial conditions, these spins move in unison and therefore the self-consistent ``magnetic field'' (or superconducting order parameter in the language of BCS theory) is  periodic in time. The reduced BCS model is integrable and there exists a very elegant prescription for constructing exact non-equilibrium solutions to it, developed primarily by Yuzbashyan and collaborators.~\cite{Emil1,classify,spectroscopy} These solutions contain, in particular, exact spin dynamics in a periodic time-dependent field that can be expressed in terms of elliptic functions. In this paper, we generalize such anomalous soliton solutions of Yuzbashyan \cite{asol} to encompass a wider range of time dependencies relevant to the problem of TLS dynamics, which is of our primary interest.

This paper is organized as follows: Sec.~II summarizes a general  mathematical structure behind the ``reverse engineering'' approach to constructing exact solutions for non-linear TLS dynamics. The specific Ansatz and technical details of our particular family of solutions for periodically-driven TLS dynamics are given in Sec.~III. In Sec.~IV, we use some representative solutions to illustrate the emergence of the coherent destruction of tunneling phenomenon. We also derive the spectrum of exact dielectric
 response function due to an ensemble of identical charge TLS in the presence of dissipation, which is introduced phenomenologically. In Sec. V we provide a summary of our results. In the Appendices we list  some technical details of our calculations as well as useful relations aimed to shed more light on the subtle features of our theory.

\section{General Framework for Constructing Exact Solutions}

In this paper, we derive a family of exact solutions for the non-dissipative TLS dynamics subject to an external ac-field. The main ingredient of our approach is a special Ansatz for the TLS's dynamics that  corresponds to periodic-in-time but non-monochromatic external fields. Before proceeding to the specific Ansatz, let us first introduce a general algebraic framework for ``reverse engineering'' of exact solutions. We are interested in solving the non-equilibirum Schr{\"o}dinger equation for the spinor $\Psi(t)$
\beg\label{schrodinger}
i{\partial_t}\Psi(t)=\hat{\cal H}(t) \Psi(t), \quad \Psi(t)=\left(\begin{matrix} \psi_{+} \\ \psi_{-}
\end{matrix}\right).
\en
where the Hamiltonian is $\hat{\cal H}(t) = {1 \over 2} {\bf b}(t)\cdot{\hat{\bm \sigma}}$.  As mentioned in the introduction, instead of solving Eq.~(\ref{schrodinger}) for the wave-function, we can consider the Schr{\"o}dinger equation for the evolution operator that relates the initial and final states, $\Psi(t) = \hat{S}(t) \Psi(0)$.  This equation for the $S$-matrix has the form identical to Eq.~(\ref{schrodinger}):
 \beg\label{Sschrodinger}
i{\partial_t}\hat{S}(t)=\hat{\cal H}(t) \hat{S}(t), \mbox{ and } \hat{S}(0) =\hat{1}
\en
but now it is an equation for the matrix function $\hat{S}(t)$, which belongs to the two-dimensional representation of the $SU(2)$ group, while the Hamiltonian expressed in terms of $SU(2)_2$ generators belongs to the two-dimensional representation of the $\mathfrak{su}(2)$ algebra. Note that the form of Eq.~(\ref{Sschrodinger}) is such that it may be generalized to an arbitrary spin or equivalently to an arbitrary-dimensional representation of $SU(2)$ or it can be viewed as an equation of motion in the abstract group such that $\hat{\cal H}_{\rm abs}(t) = {\bf b}(t) \cdot \hat {\bm J}_{\rm abs} \in \mathfrak{su}(2)$ and $\hat{S}_{\rm abs}(t) = \exp\left[ -i {\bm \Phi}(t) \cdot \hat{\bf J}_{\rm abs} \right] \in SU(2)$, where $\hat{\bf J}_{\rm abs}$ are the corresponding generators. Therefore, a solution of the problem in a particular representation, {\em i.e.}, an explicit form of ${\bm \Phi}(t)$, immediately gives the corresponding solutions in all other representations ({\em e.g.}, a two-level-system dynamics uniquely determines a ``$d$-level system'' dynamics in the same field). This TLS problem that we are interested in corresponds to the two-dimensional generators $\hat{J_\alpha}^{(2)} = {1 \over 2} \hat{\sigma}_\alpha$ with $\hat{\sigma}_\alpha$ ($\alpha=x,y,z$) being the familiar Pauli matrices.

The problem of determining the solution, ${\bm \Phi}(t)$, from the magnetic field time-dependence ${\bf b}(t)$ is a complicated one, but the inverse problem is almost trivial. Indeed, if we select a specific $S$-matrix (defined uniquely  by the choice of a specific function, ${\bm \Phi}(t)$), the Hamiltonian will read
 \beg\label{Inv}
\hat{\cal H}(t) = i{\partial_t}\hat{S}(t)\hat{S}^\dagger(t),
\en
where
\beg\label{Sexp}
\hat{S}(t) = \exp\left[ -{i \over 2} {\bm \Phi}(t) \cdot \hat{\bm \sigma} \right].
\en
Using the algebraic identities for the Pauli matrices, we obtain the corresponding magnetic field
 \beg\label{b}
{\bf b}(t) = \dot{\Phi}\, {\bf n} + \sin{\Phi}\, \dot{\bf n} + \left(1 - \cos{\Phi} \right) \left[ {\bf n} \times \dot{\bf n} \right],
\en
where ${\bm \Phi}(t) = \Phi(t) {\bf n}(t)$, with $\left|{\bf n}(t) \right| \equiv 1$. Note that one can generate exactly-solvable
models by simply picking an arbitrary ${\bm \Phi}(t)$ dependence and using Eq.~(\ref{Inv}) to find the corresponding Hamiltonian.
However, without guidance or luck, such an analysis would generally produce complicated non-equilibrium fields that have little to do with an underlying physical problem. Let us however mention here that this procedure may be of interest to quantum computing in general, because the time-evolution governed by an $S$-matrix can be viewed as a ``gate operation'' on the spin (if the TLS/spin corresponds to a qubit rather than to a defect within a qubit). By picking ``trajectories,'' ${\bm \Phi}(t)$, on the algebra that start in the origin, {\em i.e.} ${\bm \Phi}(0)={\bf 0}$, but end at a specific point at a time $T$, one can immediately determine the non-equilibrium magnetic pulse, ${\bf b}(t)$, or a class of such pulses, that will give rise to a desired gate operator $\hat{G} \equiv \hat{S}(T) = \exp\left[ -{i \over 2} {\bm \Phi}(T) \cdot \hat{\bm \sigma} \right]$.

Let us note here that the function, ${\bm \Phi}(t)$, contains complete information about the solution to the original problem, Eq.~(\ref{schrodinger}), including the
overall quantum phase accumulated by the wave-function during the time evolution (as we shall see below, this phase is of particular interest to the  problem of dielectric response of TLSs in superconducting qubits). An interesting question is whether and how this purely quantum phase can be  restored from a solution of the corresponding classical Bloch equations that are usually considered in this context. Let us recall that a classical mapping can be achieved by introducing the
average magnetic moment,
\begin{equation}
\label{spins}
{\bf m}(t) = \Psi^\dagger(t) {\hat {\bm \sigma} \over 2} \Psi(t).
\end{equation}
Therefore, ${\bf m}^2(t) \equiv 1/4$ and the classical equations of motion for the spin moment follow from  $\partial_t {\bf m}(t) = {1 \over 2} \Psi^\dagger(t) \left[ \hat{\cal H}(t),{\hat{\bm \sigma} } \right] \Psi(t)$ and yield the familiar result
 \beg\label{Blochm}
\partial_t {\bf m}(t) = {\bf b}(t) \times {\bf m}(t).
\en
Let us recall that these Bloch equations are a saddle point of quantum spin dynamics, much in the same way that  Newton's equations of motion governed by the force, $\left[-{\bm \nabla} V({\bf r})\right]$, represent a saddle point of the action describing a quantum particle in the potential, $V({\bf r})$, and therefore do not contain direct information about quantum interference and tunneling effects. Similarly, Eqs.~(\ref{Blochm}) do not directly contain the quantum phase and to determine it one has to go back to the Schr{\"o}dinger equation. Another more abstract way to see this is by noticing that Eqs.~(\ref{Blochm}) describe the motion on a two-dimensional (Bloch) sphere, ${\bf m}(t) \in S^2$, while the original quantum problem Eq.~(\ref{Sschrodinger}) describes motion on a three-dimensional sphere since $\hat{S}_{\rm abs} (t) \in SU(2) \sim S^3$. Now let us recall that there exists the Hopf fibration such that $SU(2)/U(1) = S^2$, which summarizes the fact that classical equations, namely Eqs.~(\ref{Blochm}), represent quantum motion modulo the $U(1)$ phase dynamics. Fortunately, this phase dynamics can  generally be restored from exact dependence of the  ${\bf m}(t)$ solution, albeit in a non-local way. To see this, we can write the magnetization in terms of the $S$-matrix as follows
${\bf m}(t) ={1 \over 2} \Psi^\dagger(0)\left[\hat{S}^\dagger(t) \hat {\bm \sigma} \hat{S}^\dagger(t)\right] \Psi(0)$, where $\Psi(0)$ and the corresponding ${\bf m}(0) = \Psi^\dagger(0) {\hat {\bm \sigma} \over 2} \Psi(0)$ are initial conditions for the wave-function and Bloch magnetization, correspondingly. Using again the well-known identities for the Pauli matrices, we find the evolution matrix for the Bloch equations, ${ m}_\alpha (t) = R_{\alpha \beta}(t) { m}_\beta(0)$, as follows
 \beg\label{SBloch}
 R_{\alpha \beta} (t) = \delta_{\alpha \beta} \cos{\Phi} +
n_\alpha n_\beta \left(1 - \cos{\phi} \right) - \varepsilon_{\alpha \beta \gamma} n_\gamma \sin{\Phi}.
\en
This three-dimensional matrix describes a rotation, $\hat{R}(t) \in SO(3)$, and can be represented equivalently as
\beg\label{Rexp}
\hat{R}(t) = \exp\left[ - {\bm \Phi}(t)\cdot{\hat{\bf L}} \right], \mbox{ where } \hat{\bf L} = \Biggl(
\begin{array}{ccc}
0 & -{\bf e}_z & {\bf e}_y\\
{\bf e}_z & 0 & -{\bf e}_x\\
-{\bf e}_y & {\bf e}_x & 0\\
\end{array}
\Biggr),
\en
where $\hat{\bf L} \in \mathfrak{so}(3) \sim  \mathfrak{su}(2)$ belong to the three-dimensional vector  representation of the $\mathfrak{su}(2)$ algebra. They are related to the ``usual'' spin-$1$ representation (where $\hat{J}_z^{(3)}$ is diagonal) via a simple linear transform. 

Therefore, we see that if we know an arbitrary solution to the Bloch equations, ${\bf m}(t)$ we can at least in principle restore the function, ${\bm \Phi}(t)$,
[see, Eqs.~(\ref{Rexp}) and (\ref{Sexp})], which uniquely determines the entire quantum solution. It also suggests that if we choose an arbitrary dynamic function on a sphere we may be able to restore the quantum Hamiltonian that would give rise to it, via mappings ${\bf m}(t) \to \hat{R}(t) \to {\bm \Phi}(t) \to \hat{S}(t) \to \hat{\cal H}$. However, the second step in this chain of transforms involves effectively calculating a logarithm of the rotation matrix, which due to a complicated ``analytic'' structure of this matrix-logarithm function requires a careful calculation non-local in time.

The subsequent Sections are devoted to constructing exactly solvable periodic-in-time Hamiltonians based on a specific Ansatz for the classical
 Bloch ``magnetization,'' ${\bf m}(t)$. It further involves a restoration of the corresponding quantum $U(1)$ phase via a straightforward integration.
 More specifically, we ``reverse engineer'' the following Hamiltonian
\beg\label{Eq1}
\hat{\cal H}=\Delta_t \hat{\sigma}_x+f(t)\hat{\sigma}_z.
\en
where $f(t) = f(t + T_f)$ is a periodic function, with an arbitrary period, $T_f$. Our solution below also allows tuning of the average splitting, $\varepsilon = \left\langle f(t) \right\rangle_{T_f}$, and the AC field amplitude, ${\cal A}_f \sim \sqrt{\left\langle \left[ f(t) - \varepsilon \right]^2 \right\rangle_T}$. As mentioned in the introduction, this problem is of great importance to the physical problem of externally-driven TLS dynamics in superconducting qubits (with $\Delta_t$ corresponding to tunneling between the wells, $\varepsilon$ to a splitting of energy levels in a double-well potential, and $T_f$ and $A_f$ being the period and the amplitude of the AC-electric field correspondingly).

Our ``guess'' for the relevant Ansatz for the Bloch ``magnetization,'' ${\bf m}(t)$, is  based on a set of formal solutions  discovered in the related problem of quenched dynamics of  fermionic superfluids.~\cite{Levitov2004,Emil1,classify,spectroscopy,asol} Formally, the quenched dynamics of each individual Cooper pair is described by the Bogoliubov-de Gennes Hamiltonian, which is essentially a spin Hamiltonian that reduces to (\ref{Eq1}) after the unitary transformation $\hat{\sigma}_x\to\hat{\sigma}_z$ and $\hat{\sigma}_z\to-\hat{\sigma}_x$, with $\Delta_t$  corresponding to a single particle energy level and $f(t)$ to the superfluid order parameter. A realization of each particular form of the superfluid order parameter dynamics in a steady state
can be unambiguously determined by the initial conditions~\cite{classify} using
the exact integrability of BCS model.~\cite{Emil1} Note that a self-consistency condition for the order parameter provides a limitation on the set of functions for which the corresponding problem  is integrable and for some initial conditions periodic-in-time self-consistent dynamics, $f(t)$, can be realized. While in our TLS problem, there is no natural self-consistency constraint, such insights and constraints from the BCS problem help us narrow down the range of possible Ansatze to restore reasonable physical Hamiltonians, which are also exactly solvable by construction.

In what follows, we generalize the soliton analysis of Yuzbashyan~\cite{asol} and find a general soliton configuration, characterized by three
independent parameters, which we denote as $\Delta_\pm$ and $\Delta_a$. For the physical problem of interest, this conveniently implies that some, generally speaking, non-trivial combination of these parameters will determine the arbitrary frequency, amplitude, and the dc-component of the field. Due to the periodicity, we can generally represent the AC-perturbation as a Fourier series
\beg\label{expft}
f(t)=\varepsilon+{\cal A}_f\sum\limits_{n=1}^{\infty}\tilde{f}_n\cos(n\omega_ft).
\en
Note that for certain specific choices of the parameters $\Delta_{\pm,a}$, the leading coefficient $\tilde{f}_1\gg\tilde{f}_n$ $(n=2,3,...)$ and one recovers the limit of a monochromatic AC-field, albeit in the regime of weak driving (${\cal A}_f\tilde{f}_1\ll\textrm{max}\{\Delta_{\rm t},\varepsilon\}$). Therefore, our non-linear
analysis contains the standard linear response results as a simple special case.


\section{Non-dissipative dynamics of the ac-driven TLS}
In this Section we provide the details on the derivation of the exact solution for the TLS dynamics.
We devote the special attention to the analysis of the U(1) phase of the wave function. We also elucidate the relations between the parameters of our solution and the amplitude, phase and the dc-component of the
external field, which may be useful for experimental applications of our theory.
\subsection{Ansatz}
We now focus on the Schr{\"o}dinger equation for the half-integer spin in the magnetic field, ${\bf b}(t) = 2 (\Delta_{\rm t}, 0, f(t))$. When written in terms of spinor components, it has the form
\begin{equation}
\label{BdG}
\left\{
\begin{array}{ll}
i\dot{\psi}_{+}=\Delta_{\rm t} \psi_{-}+f(t)\psi_{+}\\
\quad i\dot{\psi}_{-}=\Delta_{\rm t}\psi_{+}-f(t)\psi_{-}
\end{array}
\right..
\end{equation}
The corresponding Bloch equation is
\beg\label{Bloch}
\dot{\bf m}(t)=2(\Delta_{\rm t},0,f(t)) \times{\bf m}(t).
\en
Let us now make the following Ansatz for its exact solution:~\cite{asol}
\beg\label{ansatz}
m_x=D-Cf^2, \quad m_y=B\dot{f}, \quad m_z=Af(t)+F.
\en
From two of the Eqs. (\ref{Bloch}) we find $A=2\Delta_{\rm t} B$ and $B=C$.
Thus among five parameters in (\ref{ansatz}) only three are independent: $F, B$ and $D$.
The equation for the external field, $f(t)$, can be obtained from (\ref{ansatz}) using the
condition ${\bf m}^{2}=1/4$. This resulting equation for the function $f(t)$ acquires the form
\beg\label{fdot}
\dot{f}^2=-f^4-4c_2f^2+8c_1f-4c_3,
\en
where coefficients $c_j$ are given by some combinations of parameters $B, D$ and $F$ [see Eqs.~(\ref{BD}) below]. Equation (\ref{fdot}) can be cast to  a more symmetric form, using another set of parameters $\Delta_a$ and $\Delta_{\pm}$, which are chosen to be positive and are related to coefficients $c_j$ as follows:
\beg\label{params}
\begin{split}
&c_1=-\frac{\Delta_a}{4}(\Delta_{+}^2-\Delta_{-}^2), \\
&c_2=-\frac{1}{4}(\Delta_{+}^2+\Delta_{-}^2+2\Delta_a^2), \\
&c_3=-\frac{1}{4}(\Delta_{+}^2-\Delta_a^2)(\Delta_{a}^2-\Delta_{-}^2).
\end{split}
\en
Without loss of generality and to be more specific we also assume $\Delta_{+}\geq\Delta_{-}$
for the remainder of this paper, while $\Delta_a$ can be assigned an arbitrary value.
By virtue of expressions (\ref{params}) equation (\ref{fdot}) now reads
\beg\label{fdot2}
\dot{f}^2=[(f-\Delta_{a})^2-\Delta_-^2][\Delta_{+}^2-(f+\Delta_a)^2].
\en
Below we will make several transformations that allow us to reduce (\ref{fdot2}) to an equation
for the Weierstrass elliptic function.~\cite{Tables} Firstly, let us introduce a function, $y(t)$,
\beg\label{yt}
f(t)=\Delta_{+}\left[\frac{2}{y(t)}-1\right]-\Delta_a
\en
which satisfies the following equation
\beg\label{ydot}
\left(\frac{dy}{dx}\right)^2=4(y-a_{+})(y-a_{-})(y-1), \quad x=\frac{\Delta_{+}t}{\sqrt{a_{+}a_{-}}},
\en
where $a_{\pm}=2\Delta_{+}/(\Delta_{+}+2\Delta_{a}\pm\Delta_{-})$. Now, Eq.~(\ref{ydot}) can be easily reduced to
a well-known equation for the Weierstrass elliptic function by rescaling the parameters via the transformation
\beg\label{Zxyx}
y(x)=Z(x)+\frac{a_{+}+a_{-}+1}{3},
\en
so that
\beg\label{Zx}
\left(\frac{dZ}{dx}\right)^2=4(Z-e_1)(Z-e_2)(Z-e_3),
\en
where parameters $e_j$ satisfy the following conditions $e_1>e_2>e_3$ and $e_1+e_2+e_3=0$.
Coefficients $e_j$ are determined by the parameters $\Delta_a$ and $\Delta_{\pm}$. The specific expressions for the coefficients $e_j$,
however, depend on the relative values of the initially introduced set of parameters
and are given in Appendix A. Solution of the equation (\ref{Zx}) is
\beg\label{Weier}
Z(x)={\cal P}(x+\omega'), \quad \omega'=\frac{{\mathbf K}(\kappa')}{\sqrt{e_1-e_3}},
\en
where ${\cal P}(x)$ is a Weierstrass elliptic function, ${\mathbf K}$ is a complete elliptic integral of the first kind~\cite{Tables} and
$\kappa'=\sqrt{(e_1-e_2)/(e_1-e_3)}$. Function $Z(x)$ is a doubly-periodic function with the period along the physical time axis determined by, $l=2\omega$,
where $\omega={\mathbf K}(\kappa)/\sqrt{e_1-e_3}$ and $\kappa=\sqrt{1-\kappa'^2}$ is a modulus of elliptic functions. Combining (\ref{Weier}) with Eqs.~(\ref{Zxyx}) and (\ref{yt}) allows us to
express $f(t)$ in terms of elliptic functions. Expression for $f(t)$ can be compactly written
in terms of Jacobi elliptic functions. Just as it is the case for the parameters $e_j$, the particular form of the resulting expression depends on the relation between $\Delta_a$ and $\Delta_{\pm}$ (see Appendix A).

\begin{figure}[h]
\includegraphics[scale=0.27,angle=0]{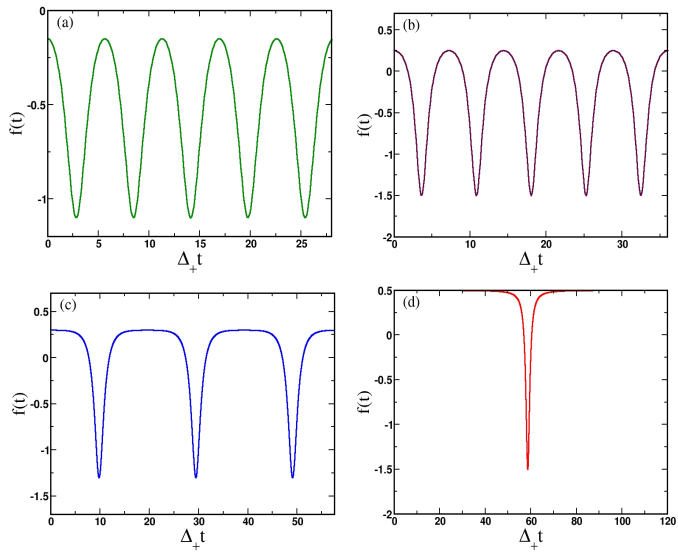}
\caption{Plots of the function $f(t)$ (\ref{ftexact}) in units of $\Delta_{+}$
for different values $\Delta_a$:
(a) $\Delta_a=0.1\Delta_{+}$, $\Delta_{-}=0.3\Delta_{+}$;
(b) $\Delta_a=0.5\Delta_{+}$, $\Delta_{-}=0.3\Delta_{+}$; (c) $\Delta_a=0.3\Delta_{+}$,
$\Delta_{-}=0.1\Delta_{+}$; (d) $\Delta_a=0.5\Delta_{+}$,
$\Delta_{-}=0.001\Delta_{+}$. We note that for the choice of the parameters (d) the period
of $f(t)$ diverges. The curves above are plotted for the value of $\Delta_{\rm t}=0.5\Delta_{+}$.}
\label{Fig2}
\end{figure}

All cases considered here are summarized by the following compact expression for the function, $f(t)$, written in terms of Jacobi elliptic function $\textrm{sn}$ as
follows:
\beg\label{ftexact}
f(t)=\Delta_{+}\frac{\eta_{+}\text{sn}^2(z,\kappa)-1}{\eta_{-}\text{sn}^2(z,\kappa)+1}-\Delta_a,
\en
where variable $z$ is
\beg\label{ut}
z=\frac{(t-t_0)}{2}\sqrt{\left[(\Delta_{+}+2\Delta_a)^2-\Delta_{-}^2\right](e_1-e_3)}
\en
and $t_0=-\omega'\sqrt{a_{+}a_{-}}/\Delta_{+}$.
If we consider $\Delta_{\pm}$ fixed, then the parameters $\eta_\pm$ are given by one of the following expressions depending on the
value of $\Delta_a$:
\beg\label{etas}
\begin{split}
\eta_\pm=\left\{\begin{array}{cc}
\frac{1}{e_1-e_3}\pm 1, & \Delta_a>\frac{\Delta_{+}+\Delta_{-}}{2} \\
\frac{1}{e_1-e_3}\pm\kappa^2, & \frac{\Delta_{+}-\Delta_{-}}{2}\leq\Delta_a\leq\frac{\Delta_{+}+\Delta_{-}}{2} \\
\frac{1}{e_1-e_3}, & \Delta_a<\frac{\Delta_{+}-\Delta_{-}}{2}
\end{array}\right..
\end{split}
\en
Fig.~2 displays some representative dependencies of the driving field from the class of solutions described by Eqs.~(\ref{ftexact}), (\ref{ut}), and (\ref{etas}). Note that the curve in Fig.~2a is visually indistinguishable from a harmonic periodic signal, Fig.~1b and Fig.~1c contain apparent non-monochromatic contributions to the periodic signal, and finally Fig.~2d provides an example of a degenerate case, or a single soliton, where the period of the elliptic  function is taken to be infinite. Fig.~3 shows dynamic trajectories of the ``magnetization'' on the Bloch sphere given by exact Eq.~(\ref{ansatz})  that correspond to these particular  $f(t)$-dependencies.

\begin{figure}[h]
\includegraphics[scale=0.27,angle=0]{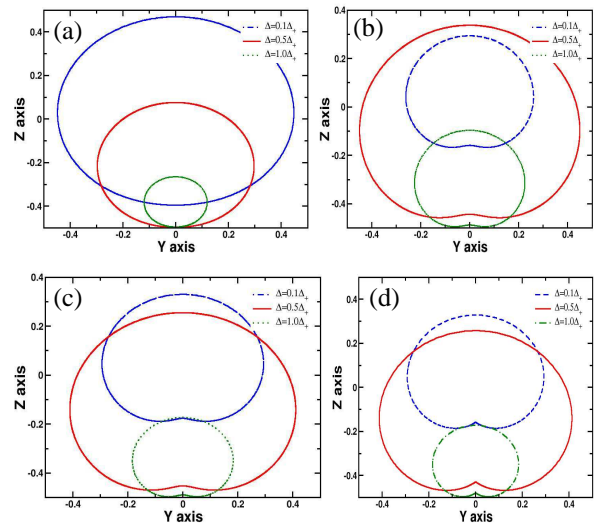}
\caption{TLS dynamics on the Bloch sphere (\ref{Bloch},\ref{ansatz}). Trajectories of TLS
for the solutions described by Eq. (\ref{ftexact}) and depicted in Fig.~2 for the various set of parameters
$\Delta_a$ and $\Delta_\pm$. The latter take the same values used on Fig.~2.}
\label{Fig1b}
\end{figure}
\begin{figure}[h]
\includegraphics[scale=0.32,angle=0]{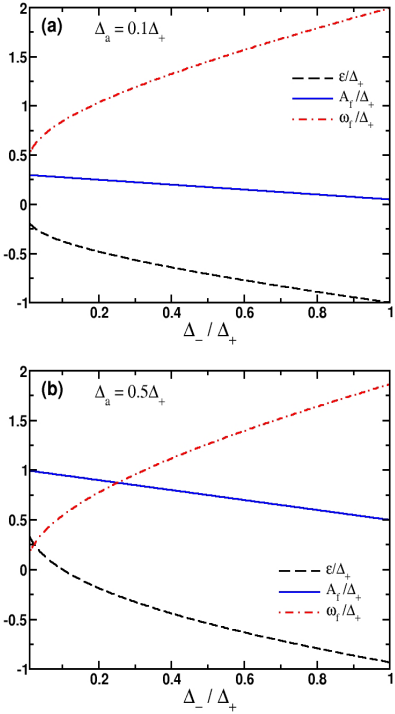}
\caption{Plots of the amplitude ${\cal A}_f$, frequency $\omega_f$ and dc-component
$\varepsilon$ of the external field $f(t)$, Eq. (\ref{expft}): (a) $\Delta_a=0.1\Delta_{+}$, $\Delta_{\rm t}=0.3\Delta_{+}$;
(b) $\Delta_a=0.5\Delta_{+}$, $\Delta_{\rm t}=0.3\Delta_{+}$.}
\label{Fig3}
\end{figure}

From the expression for the external field (\ref{ftexact}) it is, however, not immediately clear
what set of parameters correspond to the regimes of weak and strong ac-driving.
To clarify this issue, let us re-write (\ref{ftexact}) in the form more useful for practical applications.
Let us first explicitly derive the amplitude, frequency and the dc-component of function $f(t)$.
The period and the amplitude of oscillations of  $f(t)$ can be immediately deduced from (\ref{ftexact},\ref{ut}):
\beg\label{periodAmp}
\begin{split}
&T_f=\frac{4{\mathbf K}(\kappa)}{\sqrt{\left[(\Delta_{+}+2\Delta_a)^2-\Delta_{-}^2\right](e_1-e_3)}}, \\
&{\cal A}_f=\frac{\Delta_{+}}{2}\left(\frac{\eta_{+}+\eta_{-}}{\eta_{-}+1}\right).
\end{split}
\en
Lastly, the average value of the function $f(t)$ over its period is
\beg\label{ftav}
\begin{split}
\langle f(t)\rangle=&\frac{\Delta_{+}\eta_{+}}{\eta_{-}}\left[1-\frac{(\eta_{+}+\eta_{-})}{\eta_{+}{\mathbf K}(\kappa)}\Pi(-\eta_{-},\kappa)\right]-\Delta_a\equiv\varepsilon,
\end{split}
\en
with ${\mathbf K}(\kappa)$ and $\Pi(n,\kappa)$ being an complete elliptic integral of the first and third kind correspondingly. As we have already mentioned,
quantity (\ref{ftav}) describes the dc-component of the external field. One can view Eqs. (\ref{periodAmp},
\ref{ftav}) as the definition of yet another set of parameters ${\cal A}_f$, $\omega_f=2\pi/T_f$ and $\varepsilon=\langle f(t)\rangle$, which allows us to cast external field $f(t)$ into the form given by
(\ref{expft}). We plot the dependence of these parameters on the ratio
$\Delta_{-}/\Delta_{+}$ in Fig.~\ref{Fig3} for different values of $\Delta_a$ while keeping the value of
 $\Delta_{\rm t}$ fixed. As we can see from \ref{Fig3} the limits of strong and weak ac-driving are easily attainable
 with the frame of our solution. In particular, we see that the regime of the strong ac-driving should be
 achieved for moderate values of $\Delta_a$ and $\Delta_{-}/\Delta_{+}\sim 0.2$.

Expressions (\ref{ftexact},\ref{ut},\ref{etas}) constitute one of the main technical results of this paper. To get a further insight into the properties of our solution we refer the reader to Appendix B where we consider few limiting cases for the function (\ref{ftexact}). Quite generally, our solution represents the superposition of monochromatic waves with frequencies integer multiples of $\omega_f=2\pi/T_f$. As discussed in the Appendix B, solution (\ref{ftexact}) can be reduced to the monochromatic wave with frequency
$2\Delta_{+}$ when $\Delta_a=0$ and $\Delta_{-}\simeq\Delta_{+}$.

\subsection{Wave function}
Having determined the form of the periodic field $f(t)$ we employ the relations (\ref{spins}) to compute the
amplitudes $\psi_+(t)$ and $\psi_-(t)$. First let us represent these functions as follows~\cite{spectroscopy}
\beg\label{phases}
\psi_{\pm}(t)=|\psi_{\pm}(t)|e^{\mp i\phi(t)}e^{i\alpha(t)}.
\en
From these expressions, it follows that absolute values of the components $\psi_{+}$ and $\psi_{-}$ as well as their relative phase $\phi(t)$ are determined by the instantaneous value of magnetization (\ref{ansatz}). From
Eqs. (\ref{spins},\ref{ansatz}), we find
\beg\label{moduv}
\begin{split}
|\psi_{\pm}(t)|=&\sqrt{\frac{1}{2}\pm 2\Delta_{\rm t} Bf(t)\pm F},  \\
\tan[2\phi(t)]=&\frac{\dot{f}}{(D/B)-f^2(t)},
\end{split}
\en
where parameters $B$ and $D$ are determined from
\beg\label{BD}
\begin{split}
&\frac{D}{B}=2(\Delta_{\rm t}^2-c_2), ~B=\frac{1}{4\sqrt{(\Delta_{\rm t}^2-c_2)^2+\frac{c_1^2}{\Delta_{\rm t}^2}-c_3}},\\
&F=-2c_1B/\Delta_{\rm t}.
\end{split}
\en
and parameters $c_j$'s are given by (\ref{params}). Note that apparent ambiguity in signs for the
parameters $B$ and $D$ as well as for parameter $F$ is resolved
by fulfilling the condition $m^2=1/4$.
\subsection{Restoring the U(1) phase}
\begin{figure}[h]
\includegraphics[scale=0.32,angle=0]{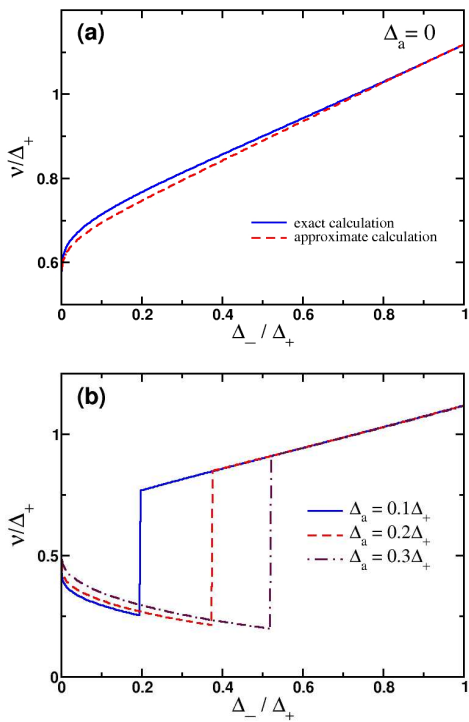}
\caption{Dependence of exponent $\nu$ as a function of the ratio $\Delta_{-}/\Delta_{+}$
for various values of $\Delta_a$. On panel (a) we compare the result of numerical computation
of $\nu$ from (\ref{dota}) and compare them with approximate expression (\ref{nuappr}) when
$\Delta_a=0$. Panel (b) shows the dependence of $\nu$ for $\Delta_a\not=0$.}
\label{Fig4}
\end{figure}

It has been mentioned above that the common phase $\alpha(t)$ has to be determined
from the solution of the equations (\ref{BdG}). At first sight the resulting equation for $\alpha(t)$ appears
to be very complicated, but it can be significantly simplified using Eqs.~(\ref{moduv}), so that
\beg\label{adotsx}
\dot{\alpha}=-\frac{1}{2}\frac{f(t)m_z(t)}{\left[{1}/{4}-m_x^2(t)\right]}.
\en
After some algebraic manipulations, we find
\beg\label{dota}
\begin{split}
&\alpha(t)=\int\limits_0^t\left\{\Delta_{\rm t}\left[\frac{d_{+}^2}{f^2(t')-d_{+}^2}-\frac{d_{-}^2}{f^2(t')-d_{-}^2}\right]+\right.\\&\left.+
\frac{F}{2B^2}\frac{f(t')}{\left[f^2(t')-d_{+}^2\right]\left[f^2(t')-d_{-}^2\right]}\right\}dt'+\alpha_0,
\end{split}
\en
where $\alpha_0$ is determined by the initial conditions,
$d_{\pm}^2=(1/2B)\pm D/B$. One can evaluate the integral in (\ref{dota}) exactly
and express in terms of elliptic $\sigma$ and $\zeta$ functions (see Apendix C for details of this calculation). We note that on the grounds of Floquet theory  we
can represent an expression for the phase $\alpha(t)$ as a sum of two terms:
\beg\label{at}
\alpha(t)-\alpha_0=-\nu t+\gamma(t),
\en
where $\gamma(t)=\gamma(t+T_f)$ is a periodic function and $\nu$ is a constant. Analytic expression for both
of these quantities can be extracted from the analytic expression for $\alpha(t)$ listed in Appendix C.
For example, from (\ref{at}) it follows $\nu=[\alpha(t)-\alpha(t+T_f)]/T_f$. In the limit when $\Delta_a=0$
and $\Delta_{+}=\Delta_{-}$ we find $\nu=(\Delta_{\rm t}^2+\Delta_{+}^2)^{1/2}$, while in the limit when
$\Delta_a=0$ and $\Delta_{-}=0$ we obtain $\nu=\Delta_{\rm t}$. For a general set of values $\Delta_a$ and
$\Delta_{\pm}$ the resulting expression for $\nu$ is not as simple as those listed above.
For practical purposes, however, one can construct an approximate expression for $\nu$.
By analyzing the behavior of $\alpha(t)$ numerically we find that for $\Delta_a=0$,
frequency $\nu$ can be approximated (see Fig. \ref{Fig4}a) by:
\beg\label{nuappr}
\nu(\Delta_a=0)\approx \frac{1}{T_f}\int\limits_0^{T_f}\sqrt{\Delta_{\rm t}^2+f^2(t)}dt.
\en
We find qualitatively different behavior of $\nu$ as a function of $\delta=\Delta_{-}/\Delta_{+}$ for nonzero
$\Delta_a$. In that case, there appears to be a discontinuity in $\nu$ at some critical value of
$\Delta_{-}/\Delta_{+}$. The source of this discontinuity at least for small $\Delta_a$
lies in the fact that $\dot{\alpha}(t) \propto m_z(t)$ changes sign
during its time evolution. For non-zero $\Delta_a$ there are always exists
$\delta_c$ such that $m_z(t=0)=0$, while for $\delta>\delta_c$ one observes $m_z(0<t_{1,2}<T_f)=0$.
The sign change in $m_z(t)$ implies that the derivative of the quantum phase will change sign also (\ref{adotsx}),
so that the subsequent integration yields the value of $\nu$ smaller than the one found
for $\delta<\delta_c$, Fig. \ref{Fig4}b.

In order to get further insight into the physical meaning of the quantity $\nu$,
we can employ the analogy between the
TLS and spin-$1/2$ and define the magnetization
${M}_\alpha(t)=\langle\Psi_g(t)|\hat{\sigma}_\alpha|\Psi_g(t)\rangle/2$, where $\Psi_g(t)$ is a general solution of the Schr\"{o}dinger equation and can be expressed as a linear combination of the particular solution $\Psi(t)$ (see below). Then one can show~\cite{spectroscopy} that the dynamics of the vector ${\vec M}$ can be represented as a linear superposition of
vector ${\vec m}(t)$ precessing with the frequency of the field $f(t)$ and a vector ${\vec h}(t)$
such that ${\vec h}\cdot{\vec m}=0$.
Each component of the latter oscillates with frequency $\nu$. Our results from Fig. 3b suggest that
the rate of precession of vector ${\vec h}$ will be significantly reduced as one tunes the parameter
$\Delta_{-}/\Delta_{+}$.

The solution of the Schr\"{o}dinger equation we described above is only a particular solution from
which the general solution can be constructed straightforwardly by taking advantage of the underlying symmetries of Eqs. (\ref{BdG}). A general solution for the wave function, $\Psi^\dagger=(\psi_{+}^*, \psi_{-}^*)$, can be presented as
\beg\label{genPsi}
\Psi_g(t)={C_1}\left(\begin{matrix} \psi_{+}(t) \\ \psi_{-}(t) \end{matrix}\right)+
{C_2}\left(\begin{matrix} {\psi}_{-}^*(t) \\ -{\psi}_{+}^*(t)
\end{matrix}\right),
\en
where $C_{1,2}$ are integration constants, which satisfy $|C_1|^2+|C_2|^2=1$ and are
to be determined from the initial conditions. For example,
for the specific choice of an initial condition when the TLS at $t=0$ resides in one of its two states,
\beg\label{psi0}
\Psi_g(0)=\left(\begin{matrix} 1 \\ 0 \end{matrix}\right),
\en
the coefficients $C_{1,2}$ are
\beg\label{C12}
C_1=\psi_{+}^*(0), \quad C_2=\psi_{-}(0).
\en

Expressions listed in this subsection amount to full description of the ac-driven
dynamics of an isolated TLS. In the next Section, we will briefly outline several
applications of our theory. For simplicity, we will mostly focus on the properties
of the non-dissipative dynamics.

\section{Experimental manifestations}
In this Section we discuss the physical behavior of several quantities which can
be probed experimentally for various physical realizations of the TLS.
Before we proceed with the discussion on the application of our results and computation of
physical observables, we derive the expression for the evolution operator and the density matrix
which will allow us to compute probabilities which characterize the dynamics of the TLS.

Evolution operator $\hat{S}(t)$ is defined by
\beg\label{defSt}
\Psi_g(t)=\hat{S}(t)\Psi_g(0).
\en
From expressions (\ref{genPsi}) one can always write down a general expression
for the evolution operator, which is valid for  arbitrary initial conditions:
\beg\label{St}
\begin{split}
\hat{S}(t)=
&\left(\begin{matrix}
\psi_{+}(t) & \psi_{-}^*(t) \\
\psi_{-}(t) & -\psi_{+}^*(t)
\end{matrix}
\right)
\left(
\begin{matrix}
\psi_{+}^*(0) & \psi_{-}^*(0) \\
\psi_{-}(0) & -\psi_{+}(0)
\end{matrix}
\right)
\end{split}
\en
Note that it is now straightforward to derive the density matrix from (\ref{St}) using the following expression:~\cite{Feynman}
\beg\label{rho}
\hat{\rho}(t)=\hat{S}(t)\hat{\rho}_0\hat{S}^{\dagger}(t),
\en
where $\hat{\rho}_0$ is the density matrix of an initial state of the TLS.
The expressions (\ref{St},\ref{rho}) can be used as a basis to analyzed the effects of the environment dissipation on the dynamics of the TLS. In particular, one can determine  the probability of the TLS to remain in the initially prepared state $P_{{\uparrow}\to\uparrow}(t)$.

\subsection{Coherent destruction of tunneling}
The phenomenon of the coherent destruction of tunneling (CDT) has been predicted theoretically~\cite{CDT1,CDT2,CDT3} for various physical realizations. Qualitatively, this phenomenon can be interpreted
as the dynamical trapping of the TLS in one its states. For example, CDT occurs when the survival probability of the initial state dynamically approaches unity. This phenomenon has its counterpart known
in literature as driving induced tunneling oscillations. This effect
has been first analyzed theoretically in a series
of papers~\cite{DITO1,DITO2,DITO3} and observed experimentally for the
first time by Nakamura \emph{et al.}~\cite{DITO4}
\begin{figure}[h]
\includegraphics[scale=0.32,angle=0]{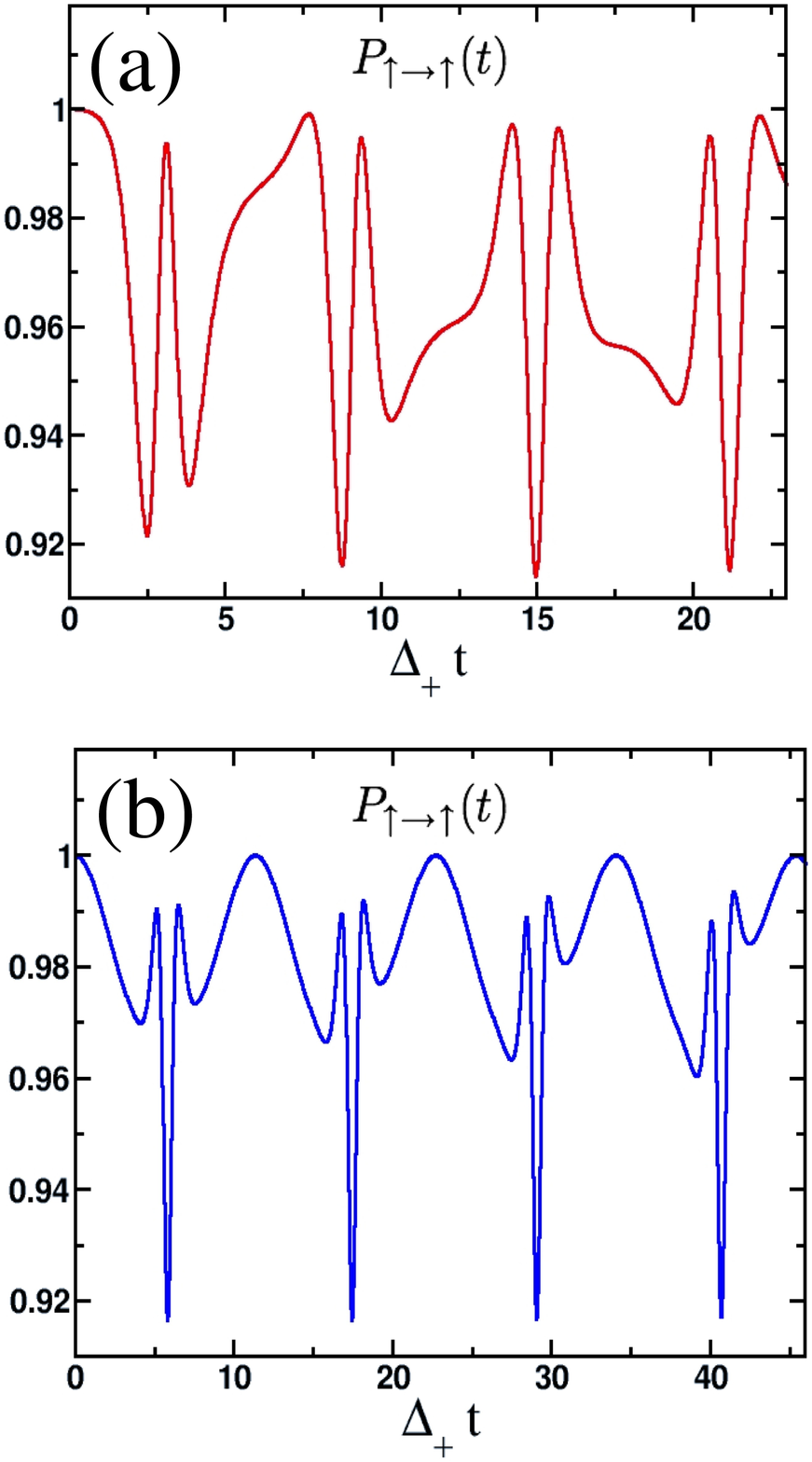}
\caption{ Plots of the return probability $P_{\uparrow\to\uparrow}(t)$, Eq (\ref{prob}),
in the limit of the strong ac-driving: (a) $\varepsilon=8.5\Delta_{\rm t}$, ${\cal A}_f=13.5\Delta_{\rm t}$, $\omega_f=21\Delta_{\rm t}$; (b) $\varepsilon=0.1\Delta_{\rm t}$, ${\cal A}_f=13.5\Delta_{\rm t}$, $\omega_f=8\Delta_{\rm t}$.}
\label{Fig5}
\end{figure}
To compute the survival probability $P_{{\uparrow}\to\uparrow}(t)$ we can use the density matrix (\ref{rho}).
It is, however, easier to use an expression for the wave function (\ref{genPsi}) with the initial conditions
(\ref{psi0},\ref{C12}). In particular, let us choose the initial amplitudes such that both $C_1$ and $C_2$
are real and introduce an angle $\vartheta$, so that $C_1=\cos(\vartheta/2)$.

After some algebra, we obtain the following expression for the return probability
\beg\label{prob}
\begin{split}
P_{{\uparrow}\to\uparrow}(t)=&\frac{1}{2}+\cos\vartheta\cdot m_z(t)+
\\&+\sin\vartheta\cos[2\alpha(t)]\cdot \sqrt{\frac{1}{4}-m_z^2(t)}.
\end{split}
\en
The CDT occurs when $P_{{\uparrow}\to\uparrow}(t)\approx 1$ and we assume the initial conditions (\ref{psi0}). From (\ref{prob}) it follows that if we perform an averaging over time frame longer than $T_f$ and
$T_h=2\pi/\nu$, the third term in (\ref{prob}) averages out to zero, so that employing (\ref{ansatz}) we find
\beg\label{prob0}
\langle P_{\uparrow\to\uparrow}(t)\rangle\simeq \frac{1}{2}+2\cos\vartheta\Delta_{\rm t} B
\left(\varepsilon-\frac{c_1}{\Delta_{\rm t}^2}\right).
\en
This equation approximately determines the parameter range for which CDT occurs.
Fig.~5 displays representative results for the return probability
and illustrates the CDT phenomenon: As we can see, in the limit of strong driving, i.e. when ${\cal A}_f\gg\Delta_{\rm t}$ and
$\omega_f\gg\Delta_{\rm t}$, the return probability remains of order unity, which implies that the tunneling processes
become strongly suppressed. We also have found
that CDT remains robust and is present as long as the parameters $\Delta_a$ and $\Delta_{\pm}$ are such that
the dynamics of the TLSs remains in the strong driving regime. This qualitative behaviour of tunneling was found to be  essentially independent of the ratio $\varepsilon/\Delta_{\rm t}$. These our findings agree qualitatively with the results reported previously in Ref.~[\onlinecite{Grifoni2009}] for the monochromatic AC-field.  Finally, we note that if a system of charged TLSs, e.g. OH-rotors present in Al$_2$O$_3$ dielectrics, is driven into such non-linear CDT regime by an external AC electric field, then the TLS tunneling and the corresponding dipole polarization dynamics will be strongly reduced.  This suggest that a strong non-linear drive may actually correspond to lower losses.

\subsection{Dielectric response}
Fig. 1 provides a pictorial example of a TLS charged defect, -- an OH-rotor, which is one of the most likely candidates of
physical two-level-systems responsible for dielectric losses in superconducting qubits. This rotor has a non-zero dipole moment
${\vec p}$ and, therefore, responds to an applied external electric field ${\vec {\cal E}}(t)$. In the absence of other interactions which may affect TLS dynamics, the Hamiltonian describing the dynamics is (\ref{Eq1}) with $f(t)=\varepsilon+{\vec p}\cdot{\vec{\cal E}(t)}$. By construction, the average
dipole moment of an isolated TLS is determined by the following average within the spin mapping~\cite{Hunklinger1981}
\beg\label{avdip}
{\vec d}(t)=m_z(t){\vec p}.
\en
The  linear dielectric response function can be computed from (\ref{avdip}) by differentiating the
corresponding components of the average dipole moment with respect to the amplitude of an external
field ${\vec {\cal E}}$. To define a non-linear dielectric response function corresponding to a solution
 $m_z(t)$, which generally is a complicated function of the amplitude, we consider the spin-spin correlation
function. Up to a pre-factor, given by the angle between the initial direction of the dipole moment
relative to the external electric field, the dielectric response of an isolated TLS is defined by
the Fourier component of the following correlator~\cite{Hunklinger1981}
\beg\label{diel}
\epsilon(\omega)=\frac{i}{4}\int\limits_0^\infty e^{i\omega t}e^{-t/\tau}
\langle[\hat{\sigma}_{z}(t),\hat{\sigma}_{z}(0)]\rangle dt
\en
where square brackets denote a commutator between the corresponding spin operators.
The exponential prefactor describes the dissipative effects of the environment and averaging is
taken over the initial state of the TLS. We are introducing  the dissipative effects on a
phenomenological level only and ignore the difference between the relaxation and
dephasing processes. This is sufficient to get insight into the general properties of the exact spectrum
of the dielectric response due to an ensemble of identical TLS. Operators $\hat{\sigma}_\alpha(t)$ in Eq.~(\ref{diel})
correspond to the Heisenberg representation:
\beg\label{Heis}
\hat{\sigma}_{z}(t)=\hat{S}^\dagger(t)\hat{\sigma}_{z}\hat{S}(t).
\en
We remind the reader that formally the evolution operator $\hat{S}(t)$ is given by
\beg
\hat{S}(t)=\hat{T}\exp\left[-i\int\limits_0^{t}H(t')dt'\right],
\en
with $\hat{T}$ being a time-ordering operator.
\begin{figure}[h]
\includegraphics[scale=0.32,angle=0]{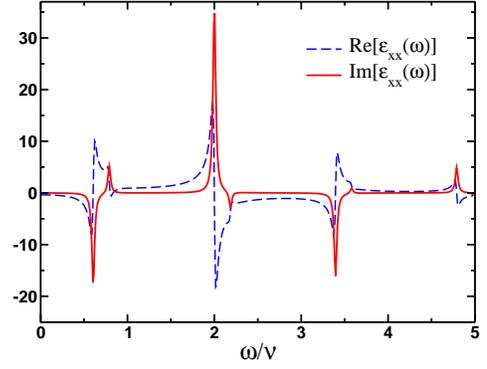}
\caption{ Plot of real and imaginary part of the response function $\epsilon(\omega)$. Note that discontinuities in real part and the peaks in imaginary part of the
response function appear at frequencies $\omega_{dis}=2(n\omega\pm\nu)$, $(n=0,\pm1,\pm2 ...)$ in agreement with expression (\ref{disc}). These plots has been obtained for the following values of the
parameters: $\Delta_a=0.15\Delta_{+}$, $\Delta_{-}=0.3\Delta_{+}$ and $\Delta_{\rm t}=0.5\Delta_{+}$. }
\label{Fig6}
\end{figure}
Using (\ref{St}) and assuming the initial conditions $\Psi^\dagger(0)=(a^*,b^*)$, for the
correlator ${\cal K}(t)=\frac{i}{4}\langle[\hat{\sigma}_{z}(t),\hat{\sigma}_{z}(0)]\rangle$ under the integral in (\ref{diel}) we find:
\beg\label{diel2}
\begin{split}
&{\cal K}(t)=4(|a|^2-|b|^2)m_\perp(t)\sin[2\alpha(t)]+8i\textrm{Im}[a^*b]\times\\&\times\{m_z(0)m_x(t)-m_x(0)m_\perp(t)\cos[2\alpha(t)]\}.
\end{split}
\en
Here $m_{\perp}(t)=\sqrt{1/4-m_x^2(t)}$ and we have fixed the initial value of the field so that
$\dot{f}(0)=0$.
The subsequent time integration yields an expression for the dielectric response function.
Analytic analysis of the response function $\epsilon(\omega)$ is hindered by the fact that
the correlation function ${\cal K}(t)$ is only a quasi-periodic function of time,
since it is expressed as a combination of two periodic functions with different periods $T_f$ and
$T_h$ (\ref{at}), so that we have to resort to numerical calculation.
In Fig. 6, we present representative plots of
the real and imaginary part of $\varepsilon(\omega)$ (\ref{diel}) for the initial conditions (\ref{psi0}).
To interpret our results, we recall that the common phase, $\alpha(t)$, can be written as a sum of
a linear-in-$t$ term plus a periodic function (\ref{at}). Since $m_\perp(t)$ is a periodic function with
the period $T_f$, we can express the corresponding terms in (\ref{diel2}) in a Fourier series.
Subsequent time integration yields a response function of the following type:
\beg\label{disc}
\epsilon(\omega)=\sum\limits_{n=-\infty}^\infty\frac{\chi_n}{\omega-2n\omega_f\pm2\nu+\frac{i}{\tau}},
\en
where $\varepsilon_n$ are the corresponding Fourier coefficients. From this expression, we see that the peaks in
the imaginary part of the response function describing the energy losses due to TLSs appear  at frequencies,
$\omega_{dis}=2(n\omega\pm\nu)$, commensurate with the driving frequency but with an overall shift determined by the quantum mechanical phase collected by a TLS over one cycle, $\nu$. Note that within the linear response theory, one typically keeps
only the lowest Fourier harmonic in the spectrum ($n=0$) and neglects all others. For the case of a monochromatic field the
Fourier component with $n=0$ is kept so that
\beg
\epsilon_{lin}(\omega)=\varepsilon_0\sum\limits_{a=\pm}\frac{1}{\omega+2a\nu+\frac{i}{\tau}},
\en
and we recover the textbook result for the dielectric response function.~\cite{Hunklinger1981} For a
fixed set of parameters, however, one would only keep the largest contribution to the imaginary part of
$\epsilon(\omega)$. For example, the imaginary part of $\epsilon(\omega)$ is the largest for
$\omega^* \simeq 2\nu$. Note also that apart from a difference in the value of the resonant frequency
(which in the regime of weak driving is given by the energy that governs a stationary time evolution of a TLS
eigenstate in the absence of any perturbations, $\nu=\sqrt{\Delta_{\rm t}^2+\varepsilon^2}$,
the response functions for the case of monochromatic field and the field given by (\ref{ftexact}) are qualitatively
the same.

In the array of non-interacting TLS, the response function must be averaged over a distribution of the
barrier heights, direction of the electric field etc. We leave the detailed analysis directly applicable to
the array of non-interacting and pairwise interacting TLS for a  future publication.

\section{Conclusions}
In this paper we presented an exact solution for the problem of AC-driven dynamics of a
generic two-level system. Our approach was based on constructing a non-linear differential equation
for the driving field, which has admitted an exact solution. The key feature of our solution for
the external field is that it is fully described by three independent parameters. We have shown that
one can interpret different nonlinear combinations of these parameters as an amplitude, frequency
and a DC-component of the field.
Being very general in nature, we believe that our results and methods can be applied to  a wide variety of
experiments ranging from NMR to the analysis of dielectric losses in amorphous materials.

This research was supported by the Intelligence Advanced Research Projects
Activity (IARPA) through the US Army Research Office award W911NF-09-1-0351.
Authors would like to thank Slava Dobrovitski and Roman Lutchyn
for discussions related to this work.

\begin{appendix}
\section{calculation of the parameters $e_j$}
In this Appendix, we provide explicit expressions for the parameters $e_j$'s, which
determine the explicit form of our exact solution for the external field (\ref{Weier}). As mentioned
in the main text, the particular expressions for these parameters, $e_j$, depend on the
relative values of $\Delta_a$ and $\Delta_{\pm}$.
For the choice corresponding to
\beg\label{choice1}
\Delta_a\geq\frac{\Delta_{-}+\Delta_{+}}{2},
\en
we have
\beg
\begin{split}
&e_1^{(1)}=\frac{1}{3}(2-a_{+}-a_{-}), \\
&e_2^{(1)}=\frac{1}{3}(2a_{-}-a_{+}-1), \\
&e_3^{(1)}=\frac{1}{3}(2a_{+}-a_{-}-1).
\end{split}
\en
In the opposite case of
\beg\label{choice2}
\frac{\Delta_{+}-\Delta_{-}}{2}\leq\Delta_a\leq\frac{\Delta_{-}+\Delta_{+}}{2},
\en
we have
\beg
\begin{split}
&e_1^{(2)}=\frac{1}{3}(2a_{-}-a_{+}-1), \\
&e_2^{(2)}=\frac{1}{3}(2-a_{-}-a_{+}), \\
&e_3^{(2)}=\frac{1}{3}(2a_{+}-a_{-}-1).
\end{split}
\en
Finally when
\beg\label{choice3}
\Delta_a\leq\frac{\Delta_{+}-\Delta_{-}}{2},
\en
we have
\beg\label{third}
\begin{split}
&e_1^{(3)}=\frac{1}{3}(2a_{-}-a_{+}-1), \\
&e_2^{(3)}=\frac{1}{3}(2a_{+}-a_{-}-1), \\
&e_3^{(3)}=\frac{1}{3}(2-a_{-}-a_{+}).
\end{split}
\en
We also remind the reader, that the coefficients $a_{\pm}$ in the above equations are given by
$a_{\pm}=2\Delta_{+}/(\Delta_{+}+2\Delta_{a}\pm\Delta_{-})$.

\section{Exact solution for the function $f(t)$: special cases}
In this Appendix, we consider a few special cases, where the exact solution given by (\ref{ftexact})
(which is generally described by three independent parameters) is reduced to a degenerate function  with simpler properties, which is characterized by two parameters only. The first case we consider corresponds to $\Delta_a\to 0$. As shown below, the choice of  $\Delta_{-}\simeq\Delta_{+}$, corresponds to an external field of the following form, $f(t)\simeq\Delta_{+}[1+q\cos(2\Delta_+ t)]$ with $q\ll 1$. Another case considered here is the limit $\Delta_{-}\to 0$, but with both $\Delta_{a}$ and $\Delta_{+}$ kept finite. In that case, $f(t)$ can be represented as a single isolated soliton.

\subsection{Limit of $\Delta_a\to 0$}
Our goal here is to  recover the limiting case for our solution corresponding to $\delta_a=0$.
It can be shown that in this limit,
\beg
\kappa=\frac{1-k'}{1+k'}, \quad k'=\delta_{-}, \quad k=\sqrt{1-k'^2}
\en
and the solution for the driving field reads:
\beg
f(t)=\Delta_{+}\left\{ \frac{2}{{\cal P}[k\Delta_{+}(t-t_0)]+1-e_3}-1\right\}.
\en
We demonstrate below that the expression in the brackets can be cast into single Jacobi elliptic function
$\text{dn}(\Delta_{+}t,k)$. For this, we use the identity
\beg
{\cal P}\left(\frac{u}{\sqrt{e_1-e_3}}\right)=e_3+(e_1-e_3)\frac{1}{\text{sn}^2(u,\kappa)},
\en
such that it gives
\beg
f(t)=\Delta_{+}\left[\frac{\text{sn}^2(u,\kappa)-(e_1-e_3)}{\text{sn}^2(u,\kappa)+(e_1-e_3)}\right]
\en
and variable $u$ equals $u=\frac{1}{2}(1+\delta_{-})\Delta_{+}t$.
This expression can be further simplified by means of the following relation between the Jacobi
elliptic functions:
\beg
\text{dn}(u_1,\kappa_1)=\frac{1-\kappa\text{sn}^2(u,\kappa)}{1+\kappa\text{sn}^2(u,\kappa)},
\en
where
\beg
u_1=(1+\kappa)u, \quad \kappa_1=\frac{2\sqrt{\kappa}}{1+\kappa}
\en
Indeed, from expressions (\ref{third}) for $\Delta_a=0$ we have
\beg
\kappa=\sqrt{\frac{e_2-e_3}{e_1-e_3}}=\frac{1-\delta_{-}}{1+\delta_{-}}=\frac{1}{e_1-e_3},
\en
so that $\kappa_1=k$ and we find:
\beg
f(t)=-\Delta_{+}\text{dn}[\Delta_{+}(t-t_0),k].
\en
Finally, when $k\to 0$ ($\Delta_{-}\to\Delta_{+}$) it follows~\cite{Tables}
that
\beg
f(t)\simeq-\Delta_{+}\left[1+q\cos(2\Delta_+ t)\right], \quad q\ll 1,
\en
We find that for the special values of parameters the line shape of the external field is given
by the cosine.
\subsection{Limit $\Delta_{-}\to 0$}
To derive an explicit  form of the driving field, $f(t)$, in this case, we work with the general solution (\ref{ftexact}).
Let us first assume that
\[
\Delta_a\leq \Delta_{+}/2.
\] T
Then, the case $\Delta_{-}=0$ corresponds to $k=1$, which in turn implies
\beg
\text{sn}(u,1)=\tanh(u), \quad u=\frac{1}{2}\lambda t
\en
and $\lambda=\sqrt{\Delta_{+}^2-4\Delta_a^2}$.
After some simple algebra, we find
\beg\label{sol1}
f=-\left[\Delta_a+\frac{\lambda^2}{2\Delta_a-\Delta_{+}\cosh(\lambda t)}\right],
\en
which up to the minus sign, is exactly the same expression as the listed in Ref.~[\onlinecite{asol}].
Finally, let us consider the parameter range  with
\beg
\Delta_a\geq\Delta_{+}/2.
\en
According to the expressions above for that case $k=0$
and $\text{sn}(u,0)=\sin(\sqrt{4\Delta_a^2-\Delta_{+}^2}t/2)$. It follows:
\beg\label{sol2}
f=-\Delta_a+
\frac{4\Delta_{+}(2\Delta_a-\Delta_{+})/(2\Delta_a+\Delta_{+})}{1-\cos\left(\sqrt{4\Delta_a^2-\Delta_{+}^2}t\right)+2\frac{2\Delta_a-\Delta_{+}}{2\Delta_a+\Delta_{+}}}.
\en
We see that when $\Delta_{-}=0$ external field has a line shape of a single pulse. Note that
our solutions (\ref{sol1},\ref{sol2}) do not contradict to our assumption of the periodicity of $f(t)$ since both
these solutions correspond to the case where the period of $f(t)$ goes to infinity.

\vspace*{0.1in}

\section{calculation of the common phase $\alpha(t)$}
In this Appendix, we outline the main steps, which allow to compute the integral (\ref{dota}) exactly. The calculation includes the following transform of the special functions involved that reduces the integrand to a form amenable for exact integration
of the Weierstrass elliptic function:~\cite{Tables}
\beg\label{tabint}
\begin{split}
\int\frac{\alpha{\cal P}(u)+\beta}{\gamma{\cal P}(u)+\delta}du=&\frac{\alpha}{\gamma}u+\frac{\alpha\delta-\beta\gamma}{\gamma\delta}\times\\&\times\left[\log\frac{\sigma(u+v)}{\sigma(u-v)}-2u\zeta(v)\right],
\end{split}
\en
where $\alpha,\beta,\gamma,\delta$ are some constants, a parameter $v$ is determined from the
derivative of the Weierstass function, ${\cal P}'(v)=-\delta/\gamma$, $\sigma(u)$, and $\zeta(u)$
are the Weierstrass elliptic sigma and zeta functions.~\cite{Tables}

The next step is to write down the function, $f(t)$,
explicitly in terms of the Weierstrass function. Combining expressions (\ref{yt},\ref{Zxyx},\ref{Weier}), we have:
\beg\label{ftweier}
f(t)=-\Delta_{+}\left[\frac{{\cal P}(x+\omega')-1-e_j}{{\cal P}(x+\omega')+1-e_j}\right]-\Delta_a
\en
where $a_{\pm}=2\Delta_{+}/(\Delta_{+}+2\Delta_{a}\pm\Delta_{-})$,
$x=\frac{\Delta_{+}t}{\sqrt{a_{+}a_{-}}}$ and $j=1,2$ or $3$ depending on the
value of $\Delta_a$ (see Appendix A). Let us now consider the first integral in (\ref{dota}):
\begin{widetext}
\beg\label{c1}
\begin{split}
\int\limits_0^t\frac{d_{+}^2}{f^2(t')-d_{+}^2}dt'=\frac{\sqrt{a_{+}a_{-}}}{2\Delta_{+}}
\int\limits_{0}^x\left\{\frac{d_{+}\left[{\cal P}(x'+\omega')+1-e_j\right]}{(\Delta_{+}+\Delta_a-d_{+}){\cal P}(x'+\omega')+(\Delta_a-d_{+})(1-e_j)-\Delta_{+}(1+e_j)}-(d_{+}\to -d_{+})
\right\}dx'
\end{split}
\en
\end{widetext}
Here the index $j$ of the coefficient $e_j$ is determined by the value of $\Delta_a$ (see Appendix A).
The remaining terms can be written
in a similar form and the corresponding integrals can be evaluated  using (\ref{tabint}), as we have done for the first one (\ref{c1}). Since the resulting expressions for the $\alpha(t)$ turn out to be too cumbersome, we do not list them here.

\end{appendix}


\begin{thebibliography}{99}

\bibitem{Weiss} U. Weiss, \emph{Quantum dissipative systems} (Wold Scientific, Singapore, 2008), 3rd Ed.

\bibitem{Leggett1987} A. J. Leggett \emph{et al.}, Rev. Mod. Phys. {\bf 59}, 1 (1987).

\bibitem{AndersonTLS} P. W. Anderson, B. Halperin and C. Varma, Phil. Mag. {\bf 25}, 1 (1972).

\bibitem{YuAndersonA15} C. C. Yu and P. W. Anderson, Phys. Rev. B {\bf 29}, 6165 (1984).

\bibitem{Hunklinger1981} S. Hunklinger and A. K. Raychaudhuri, \emph{Amorphous Solids: Low-Temperature Properties}, edited by W. A. Phillips (Springer, Berlin, 1981).

\bibitem{Peter2002} V. Lubchenko and P. G. Wolynes, J. Chem. Phys. {\bf 119} (17) 9088 (2002).

\bibitem{manipulate} M. A. Nielsen, I. L. Chuang, \emph{Quantum Computations and Quantum Information}
(Cambridge Univ. Press, Cambridge, 2002).

\bibitem{Gerardot2009} B. D. Gerardot and P. \"{O}hberg, Science {\bf 326}, 1489 (2009).

\bibitem{expJJ1} J. E. Mooij, T. P. Orlando, L. Levitov, L. Tian, C. H.
van der Wal, and S. Lloyd, Science {\bf 285}, 1036 (1999).

\bibitem{expJJ2}  C. H. van der Wal, A. C. J. ter Haar, F. K. Wilhelm, R. N. Schouten, C. J. P. M. Harmans, T. P. Orlando, S. Lloyd, and J. E. Mooij, Science {\bf 290}, 773 (2000).

\bibitem{expJJ3} I. Chiorescu, Y. Nakamura, C. J. P. M. Harmans, and
J. E. Mooij, Science {\bf 299}, 1869 (2003).

\bibitem{SlavaScience1} G. D. Fuchs \emph{et al.}, Science {\bf 326}, 1520 (2009).

\bibitem{YuPRL} J. M. Martinis \emph{et al.}, Phys. Rev. Lett. {\bf 95}, 210503 (2005).

\bibitem{Wang2009} H. Wang \emph{et al.}, \emph{pre-print} arXiv:0909.0547 [cond-mat.mes-hall] (2009).

\bibitem{Grifoni2009} J. Hausinger and M. Grifoni, \emph{pre-print} arXiv:0910.0356 [quant-ph] (2009).

\bibitem{Phillips1973} W. A. Phillips, J. of Low Temp. Phys. {\bf 11}, 757 (1973).

\bibitem{Charles2010} C. Musgrave, \emph{private communication}.

\bibitem{Osborn2010} H. Paik and K. D. Osborn, Appl. Phys. Lett. {\bf 96}, 072505 (2010).

\bibitem{Levitov2004} R. A. Barankov, L. S. Levitov, and B. Z. Spivak: Phys. Rev. Lett. {\bf 93}, 160401 (2004); R. A. Barankov and L. S. Levitov, Phys. Rev. Lett. 96, 230403 (2006).

\bibitem{Emil1} E. A. Yuzbashyan, B. L. Altshuler, V. B. Kuznetsov, V. Z. Enolskii: J. Phys. A 38, 7831 (2005); E. A. Yuzbashyan, B. L. Altshuler, and V. B. Kuznetsov: Phys. Rev. B 72, 144524 (2005);
E. A. Yuzbashyan, B. L. Altshuler, V. B. Kuznetsov, V. Z. Enolskii: Phys. Rev. B 72, 220503 (2005);
E. A. Yuzbashyan and M. Dzero, Phys. Rev. Lett. {\bf 96}, 230404 (2006).

\bibitem{classify} E. A. Yuzbashyan, O. Tsyplyatyev and B. L. Altshuler, Phys. Rev. Lett. {\bf 96}, 097005 (2006).

\bibitem{Anderson1958} P. W. Anderson, Phys. Rev. {\bf 112}, 1900 (1958).

\bibitem{VG2010} V. M. Galitski, arXiv:1003.2237v1 (2010).

\bibitem{spectroscopy} M. Dzero, E. A. Yuzbashyan, B. L. Altshuler and P. Coleman,
Phys. Rev. Lett. {\bf 99}, 160402 (2007).

\bibitem{asol} E. A. Yuzbashyan, Phys. Rev. B {\bf 78}, 184507 (2008).

\bibitem{Tables} I. S. Gradstein and I. M. Ryzhik, \emph{Tables of Integrals, Series, and Products} (Academic Press, San Diego, 1994).

\bibitem{Feynman} R. P. Feynman, \emph{Statistical Mechanics: A Set of Lectures}
(Addison-Wesley, New York, 1972).

\bibitem{CDT1} F. Grossmann, T. Dittrich, P. Jung, and P. H\"{a}nggi, Phys. Rev. Lett. {\bf 67}, 516 (1991).

\bibitem{CDT2} F. Grossmann, P. Jung, T. Dittrich, and P. H\"{a}nggi, Z. Phys. B {\bf 84}, 315 (1991).

\bibitem{CDT3} F. Grossmann and P. H\"{a}nggi, Europhys. Lett. 18, 571 (1992).

\bibitem{DITO1} L. Hartmann, M. Grifoni, and P. H\"{a}nggi, J. Chem. Phys. {\bf 109}, 2635 (1998).

\bibitem{DITO2} L. Hartmann, I. Goychuk, M. Grifoni, and P. H\"{a}nggi, Phys. Rev. E {\bf 61}, R4687 (2000).

\bibitem{DITO3} I. Goychuk and P. H\"{a}nggi, Adv. Phys. {\bf 54}, 525 (2005).

\bibitem{DITO4} Y. Nakamura, Y. A. Pashkin, and J. S. Tsai, Phys. Rev. Lett. {\bf 87}, 246601 (2001).

\end{thebibliography}
\end{document}